\documentstyle[12pt]{article}
\begin{document}
\def\l{\lambda}
\def\e{\epsilon}
\def\o{\overline}
\vspace{8cm}
{\LARGE
\begin{center}
Lie point symmetries of integrable evolution equations\\
and invariant solutions
\end{center}
}
{\large
\begin{center}
Andrei K. Svinin\footnote{e-mail: svinin@icc.ru}\\
Institute of System Dynamics and Control Theory\\
Siberian Branch of Russian Academy of Sciences\\
P.O. Box 1233, 664033 Irkutsk, Russia
\end{center}
}
\centerline{{\bf Abstract}}
\begin{quote}
An integrable hierarchies connected with linear stationary Schr\"{o}dinger
equation with energy dependent potentials (in general case) are
considered. Galilei-like and scaling invariance transformations are constructed.
A symmetry method is applied to construct invariant solutions.
\end{quote}

\newpage
{\large
\leftline{{\bf I. INTRODUCTION}}
}

Symmetries of partial differential equations are used for description of the
general set of solutions, for producing families of solutions from known exact
solutions, for construction of invariant solutions etc$^{1-3}$.

Integrable evolution equations (systems of equation),
with the group point of view,
has a remarkable property: they come in hierarchies of
commuting flows which often it is possible to describe with the help
of suitable recursion operator as
$$
{\bf u}_{t_m} = \Lambda^{m-1}{\bf K}_1,
$$
where $\Lambda$ is recursion operator and ${\bf K}_1$ is some vector field.
In many cases ${\bf K}_1[{\bf u}] = {\bf u}_x$, which presents a shift of
spatial variable $x\in{\bf R}^1$.
The fact of compatibility of these flows make possible to say about
simultaneous solution ${\bf u}(x, t_2, t_3,...)$ and investigate
their group properties.

In this article we investigate integrable evolution equations (systems of equations)
associated with stationary Schr\"{o}dinger equation with energy-dependent
potential. In this case, recursion operator $\Lambda$ is known in its
explicit form. That allows us to establish transformation property
of $\Lambda$. We construct one-parameter linear point transformation
generated by the shift of spectral parameter $\l \rightarrow \o{\l} =
\l - \e$. Particular case of these transformations is well known Galilei
transformation for Korteweg---de Vries equation. Also we construct scaling invariance
group. We next apply standard symmetry method to construct
simultaneous solutions, invariant w.r.t. Galilei-like group ${\cal G}_{N,n}$.
Corresponding ansatz involve the vector-function ${\bf U} = {\bf U}(X, T_2,..., T_{N-2},
T_N)$, where $\{X=T_1, T_2,..., T_{N-2}, T_N\}$ is the set of differential
invariants. We show that this ansatz reduce first $N$ members of hierarchy
$$
{\bf u}_{t_m} = \Lambda^{m-1}[{\bf u}]{\bf u}_x = {\bf K}_m[{\bf u}],\;\;\;
m = 1,..., N
$$
to $(N-2)$ members of that, i.e.,
$$
{\bf U}_{T_m} = {\bf K}_m[{\bf U}],\;\;\;
m = 1,..., N - 2,
$$
with additional Galilean and scaling self-similarity constraint.

The plan of this article is as follows. In Sec. II, we construct
symmetry transformation groups.  Sec. III is devoted to construction
${\cal G}_{N,n}$-invariant solutions.

\vspace{0.5cm}

{\large
\leftline{{\bf II. INTEGRABLE HIERARHIES AND SYMMETRY}}
\leftline{{\bf TRANSFORMATIONS}}
}

Let us recall some relevant notions which are useful throughout this paper.
Let $M$ be a manifold
of smooth vector-functions ${\bf u} : {\bf R}^1\rightarrow{\bf C}^n$. We
denote by $A_{{\bf u}}$, the algebra of polynomials in finite collection of
variables $u_{ix}^{(k)}$.

{\it Definition 1:}  Let ${\bf K}[{\bf u}] = \left(K_1[{\bf u}],...,
K_n[{\bf u}]\right)^T\in T_{\bf u}M$
is a vector field and $\Lambda : T_{\bf u}M\rightarrow T_{\bf u}M$
is a linear operator.
The Gateaux derivatives of ${\bf K}$ and $\Lambda$ with respect to ${\bf u}$
in the direction ${\bf X}\in T_{{\bf u}}M$ are defined through the relations
$$
{\bf K}^{\prime}[{\bf u}]({\bf X}) =
\left.\frac{\partial}{\partial \tau}\right|_{\tau = 0}
{\bf K}[{\bf u} + \tau {\bf X}],\;
\Lambda^{\prime}[{\bf u}]({\bf X}) =
\left.\frac{\partial}{\partial \tau}\right|_{\tau = 0}
\Lambda[{\bf u} + \tau {\bf X}].
$$

{\it Definition 2:} The Lie derivatives of ${\bf K}$ and $\Lambda$ in the direction
${\bf X}$ are defined, respectively, as
$$
L_{{\bf X}}{\bf K} = [{\bf X}, {\bf K}] =
{\bf K}^{\prime}({\bf X}) - {\bf X}^{\prime}({\bf K}),\;
L_{{\bf X}}\Lambda = \Lambda^{\prime}({\bf X}) - [{\bf K}^{\prime}, \Lambda].
$$

The linear space $T_{\bf u}M$ endowed with the commutator $[{\bf X}, {\bf Y}]
= L_{{\bf X}}{\bf Y}$
bears the structure of the infinite Lie algebra ${\rm Vect}$.

{\it Definition 3:} The operator $\Lambda : T_{{\bf u}}M\rightarrow T_{{\bf u}}M$
is called
hereditary if its Nijenhuis torsion vanishes $^4$, i.e.
$$
T_{\Lambda}({\bf X}, {\bf Y}) = [\Lambda {\bf X}, \Lambda {\bf Y}] -
\Lambda\left\{[\Lambda {\bf X}, {\bf Y}] +
[{\bf X}, \Lambda {\bf Y}] \right\} + \Lambda^2[{\bf X}, {\bf Y}] = 0
$$
for any ${\bf X}, {\bf Y} \in T_{\bf u}M$.

Vector fields $\Lambda^{k-1}{\bf K}_1$ span abelian subalgebra in
${\rm Vect}$ if an operator $\Lambda$ is hereditary and $L_{{\bf K}_1}\Lambda = 0$
$^5$.

It is known that, with each auxiliary linear equation
\begin{equation}
\psi_{xx}(x, \l) + \left[\sum_{i=1}^nu_i(x)(-\l)^{i-1} + (-\l)^n\right]\psi(x, \l) = 0,
\label{eq:schr}
\end{equation}
one can associate the hierarchies of commuting flows$^6$
\begin{equation}
{\bf u}_{t_m} = {\bf K}_m[{\bf u}] = \Lambda^{m-1}[{\bf u}] {\bf u}_x,\;\; {\bf u}
= (u_1,..., u_n)^T
\label{eq:hier}
\end{equation}
with the hereditary recursion operator of the form
\begin{equation}
\Lambda[{\bf u}] =
\left(
\begin{array}{ccccc}
0&0&...&0&\frac{1}{4}\partial_x^2 + u_1 + \frac{1}{2}u_{1x}\partial_x^{-1}\\[0.3cm]
-1&0&...&0&u_2 + \frac{1}{2}u_{2x}\partial_x^{-1}\\[0.3cm]
0&-1&\ddots&\vdots&u_3 + \frac{1}{2}u_{3x}\partial_x^{-1}\\[0.3cm]
\vdots&\ddots&\ddots&0&\vdots\\[0.3cm]
0&...&0&-1&u_n + \frac{1}{2}u_{nx}\partial_x^{-1}
\end{array}
\right).
\label{eq:ro}
\end{equation}
It is obvious that $L_{{\bf u}_x}\Lambda = 0$.

In what follows, we define the operator $\partial^{-1}_x$ occurring in
(\ref{eq:ro}) so that $\partial^{-1}_x(f)\in A_{{\bf u}}^0$ for any
$f = f[{\bf u}]\in {\rm Im}\;\partial_x\subset A_{\bf u}^0$,
where $A_{{\bf u}}^0\subset A_{{\bf u}}$ is a ring of differential
polynomials in the fields $u_i$ with zero constant terms.
\vspace{0.4cm}

\leftline{{\bf A. Galilei-like symmetry transformations}}

In the previous paper$^{7}$ we have proposed a construction of Galilei-like
symmetry transformations for equations (\ref{eq:hier}). To make this paper
self-contained, we gives all proofs below.

Let us define the one-parameter linear transformation of the fields $u_i$
by the shift of the spectral parameter $\l$ through the relation
$$
\sum_{i=1}^{n}u_i(x)(-\l)^{i-1} + (-\l)^n =
\sum_{i=1}^{n}\o{u}_{i}(x)(-\l+\e)^{i-1} + (-\l+\e)^n.
$$
Comparing the coefficients of the same powers of $\l$, one obtains
\begin{equation}
\begin{array}{c}
\displaystyle
u_1 = F_1(\o{{\bf u}}, \e) = \sum_{i=1}^n \o{u}_{i}\e^{i-1} + \e^n,\\[0.5cm]
u_{i+1} = F_{i+1}(\o{{\bf u}}, \e) =
\displaystyle \frac{1}{i!}\frac{\partial^iF_1(\o{{\bf u}}, \e)}{\partial \e^i},
\;\;i = 1,..., n - 1.
\end{array}
\label{eq:6}
\end{equation}
It is obvious that inverse transformation to (\ref{eq:6}) is
given by
\begin{equation}
\o{u}_1 = F_1({\bf u}, -\e),\;\;\;\o{u}_{i+1} = F_{i+1}({\bf u}, -\e),\;\;
i = 1,..., n - 1.
\label{eq:7}
\end{equation}
Eq. (\ref{eq:6}) can be rewritten in vector form
\begin{equation}
{\bf u} = {\bf F}(\o{{\bf u}}, \e) = A(\e)\o{{\bf u}} + {\bf d}(\e),
\label{bf}
\end{equation}
where $A(\e)$ is an upper diagonal matrix with units in main diagonal and
${\bf d}(\e)$ is a vector. From (\ref{eq:6}) we easy obtain
\begin{equation}
A_{ik}(\e) = C_{k-1}^{i-1}\e^{k-i},\;\;k>i,\;\;
d_i(\e) = C_n^{i-1}\e^{n-i+1}.
\label{aik}
\end{equation}
Here and in what follows the symbol $C_p^q$ denotes binomial coefficient $\left(q \atop p \right)$.

{\bf Lemma 1:} {\it The recursion operator satisfy the following identity:}
\begin{equation}
\left.\Lambda[{\bf u}]\right|_{{\bf u} = F(\o{{\bf u}}, \e)} =
A(\e)\left(\Lambda[\o{{\bf u}}] + \e\right)A^{-1}(\e)
\label{eq:9}
\end{equation}

{\it Proof:} Let us denote
$$
p(\e) = p_0 + ...+ p_{n-1}\e^{n-1} + \e^n =
\left.\left(\frac{1}{4}\partial_x^2 + u_1 + \frac{1}{2}u_{1x}\partial_x^{-1}
\right)\right|_{u_1=F_1(\o{{\bf u}}, \e)}.
$$
By virtue of Eq. (\ref{eq:6}) we can write
$$
\Lambda(\e)
\stackrel{{\rm def}}{=}
\left.\Lambda[{\bf u}]\right|_{{\bf u} = {\bf F}(\o{{\bf u}}, \e) } =
\left(
\begin{array}{ccccc}
0&0&...&0&p(\e)\\[0.3cm]
-1&0&...&0&p^{\prime}(\e)\\[0.3cm]
0&-1&\ddots&\vdots&\frac{1}{2!}p^{\prime\prime}(\e)\\[0.3cm]
\vdots&\ddots&\ddots&0&\vdots\\[0.3cm]
0&...&0&-1&\frac{1}{(n-1)!}p^{(n-1)}(\e)
\end{array}
\right).
$$
It is obvious that $\Lambda(0) = \Lambda[\o{{\bf u}}]$. Since the matrix
$A(\e)$ is nondegenerate, we can rewrite Eq. (\ref{eq:9}) as
\begin{equation}
A(\e)\left(\Lambda(0) + \e\right) = \Lambda(\e)A(\e).
\label{eq:relation}
\end{equation}
The coefficients $p_i$ are operators but it is obvious that when proving
fulfillment of relation (\ref{eq:relation}) we can operate with $p_i$
like with a numbers.

We have
$$
A(\e)\Lambda(0) =
\left(
\begin{array}{ccccc}
-\e&-\e^2&...&-\e^{n-1}&p(\e)-\e^n\\[0.3cm]
-1&-C_2^1\e&...&-C_{n-1}^1\e^{n-2}&p^{\prime}(\e)-C_n^1\e^{n-1}\\[0.3cm]
0&-1&  &-C_{n-1}^2\e^{n-3}&\frac{1}{2!}p^{\prime\prime}(\e)-C_n^2\e^{n-2}\\[0.3cm]
\vdots&\vdots&\ddots&\vdots&\vdots\\[0.3cm]
0&0&...&-1&\frac{1}{(n-1)!}p^{(n-1)}(\e)-C_n^{n-1}\e
\end{array}
\right),
$$
$$
\e A(\e) =
\left(
\begin{array}{cccccc}
\e&\e^2&\e^3&...&\e^{n-1}&\e^n\\[0.3cm]
0&\e&C_2^1\e^2&...&C_{n-2}^1\e^{n-2}&C_{n-1}^1\e^{n-1}\\[0.3cm]
0&0&\e&...&C_{n-2}^2\e^{n-3}&C_{n-1}^2\e^{n-2}\\[0.3cm]
\vdots&\vdots&\vdots&  &  &\vdots\\[0.3cm]
0&0&0&...&0&\e
\end{array}
\right),
$$
$$
\Lambda(\e)A(\e)
\left(
\begin{array}{cccccc}
0&0&0&...&0&p(\e)\\[0.3cm]
-1&-\e&-\e^2&...&-\e^{n-2}&p^{\prime}(\e)-\e^{n-1}\\[0.3cm]
0&-1&-C_2^1\e&...&-C^{1}_{n-2}\e^{n-3}&\frac{1}{2!}p^{\prime\prime}(\e)-C^{n-1}_1\e^{n-2}\\[0.3cm]
\vdots&\vdots&\vdots&   &\vdots&\vdots\\[0.3cm]
0&0&0&...&-1&\frac{1}{(n-1)!}p^{(n-1)}(\e)-C_{n-1}^{n-2}\e
\end{array}
\right).
$$
Comparing diagonal-wise matrix elements and using the relation
$$
C_{k}^{k-r} - C_{k-1}^{k-r-1} = C_{k-1}^{k-r},
$$
one can verify that (\ref{eq:relation}) is fulfilled. $\Box$

The relation (\ref{eq:9}) defines transformation law of recursion operator
$\Lambda[{\bf u}]$ w.r.t. (\ref{eq:6}), but it should be noted
that operator $\partial_x^{-1}$, occurring in $\Lambda[\o{{\bf u}}]$, now acts
on a ring of differential polynomials in the transformed fields $\o{u}_i$
and its action must be consistent with this transformation. Let
${\bf F}_{*} : A_{\bf u}\rightarrow A_{\o{{\bf u}}}$ and
${\bf F}_{*}^{-1} : A_{\o{{\bf u}}}\rightarrow A_{\bf u}$ are maps
of rings of differential polynomials generating by (\ref{eq:6}). Then
we must to require that
\begin{equation}
{\bf F}_{*}^{-1}\circ\partial_x^{-1}(f)\in A_{\bf u}^0,
\label{eq:10}
\end{equation}
for any $f[\o{{\bf u}}]\in{\rm Im}\:\partial_x\subset A_{\o{{\bf u}}}^0$.

So, when computing vector fields $\Lambda^r[\o{{\bf u}}]\o{{\bf u}}_x$
we must to take into account the condition (\ref{eq:10}). For example,
\begin{equation}
\partial_x^{-1}(\o{u}_{nx}) = \o{u}_n + n\e
\label{eq:because}
\end{equation}
because ${\bf F}_{*}^{-1}\left(\o{u}_n + n\e\right) = u_n\in A^0_{{\bf u}}$.
Taking into account (\ref{eq:because}) we have
\begin{equation}
\Lambda[\o{{\bf u}}]\o{{\bf u}}_x = K_2[\o{{\bf u}}] + \frac{n}{2}\e\o{{\bf u}}_x,
\label{eq:f1}
\end{equation}
To compute $\Lambda^2[\o{{\bf u}}]\o{{\bf u}}_x$ there is a need to calculate the
$n$th component of the vector field ${\bf K}_2[\o{{\bf u}}]$. We have
$K_2[\o{u}] = \frac{1}{4}\o{u}_{xxx} + \frac{3}{2}\o{u}\o{u}_x\in{\rm Im}
\:\partial_x$ in the case $n=1$ and $K_{2,n}[\o{{\bf u}}] = -\o{u}_{n-1,x} +
\frac{3}{2}\o{u}_n\o{u}_{nx}\in{\rm Im}\:\partial_x$ for $k\ge 2$.
Taking into account (\ref{eq:10}), we obtain
\begin{equation}
\partial_x^{-1}\left(K_2[\o{u}]\right) =
\frac{1}{4}\o{u}_{xx} + \frac{3}{4}\o{u}^2 - \frac{3}{4}\e^2,
\label{eq:f2}
\end{equation}
\begin{equation}
\partial_x^{-1}\left(K_{2,n}[\o{{\bf u}}]\right) =
-\o{u}_{n-1} + \frac{3}{4}\o{u}_n^2 +
\left(\frac{1}{2}n(n-1) - \frac{3}{4}n^2\right)\e^2
\label{eq:f3}
\end{equation}
Using (\ref{eq:f1}), (\ref{eq:f2}) and (\ref{eq:f3}) we obtain:
$$
\Lambda^2[\o{{\bf u}}]\o{{\bf u}}_x = \Lambda[\o{{\bf u}}]K_2[\o{{\bf u}}] +
\frac{n}{2}\e\Lambda[\o{{\bf u}}]\o{{\bf u}}_x
$$
$$
= {\bf K}_3[\o{{\bf u}}] +
\left(\frac{1}{2}n(n-1) - \frac{3}{4}n^2\right)\e^2\o{{\bf u}}_x +
\frac{n}{2}\e\left({\bf K}_2[\o{{\bf u}}] + \frac{n}{2}\e\o{{\bf u}}_x\right)
$$
\begin{equation}
= {\bf K}_3[\o{{\bf u}}] + \frac{n}{2}\e {\bf K}_2[\o{{\bf u}}] +
\left(\frac{1}{2}n(n-1) - \frac{1}{8}n^2\right)\e^2\o{{\bf u}}_x.
\label{eq:f4}
\end{equation}
Using lemma 1 and Eqs. (\ref{eq:f1}), (\ref{eq:f4}) we obtain evolution
equations on transformed functions $\o{u}_i(x, t_2, t_3)$
$$
\o{{\bf u}}_{t_2}=\left(\Lambda[\o{{\bf u}}] + \e\right)\o{{\bf u}}_x
$$
\begin{equation}
={\bf K}_2[\o{{\bf u}}] + \left(\frac{n}{2} + 1\right)\e\o{{\bf u}}_x,
\label{eq:ut2}
\end{equation}
\vspace{0.2cm}
$$
\o{{\bf u}}_{t_3}=\left(\Lambda[\o{{\bf u}}] + \e\right)^2\o{{\bf u}}_x
$$
$$
=\Lambda^2[\o{{\bf u}}]\o{{\bf u}}_x + 2\e\Lambda[\o{{\bf u}}]\o{{\bf u}}_x +
\e^2\o{{\bf u}}_x
$$
$$
={\bf K}_3[\o{{\bf u}}] + \frac{n}{2}\e {\bf K}_2[\o{{\bf u}}] +
\left(\frac{1}{2}n(n-1) - \frac{1}{8}n^2\right)\e^2\o{{\bf u}}_x
$$
$$
+2\e\left({\bf K}_2[\o{{\bf u}}] + \frac{n}{2}\e\o{{\bf u}}_x\right) +
\e^2\o{{\bf u}}_x
$$
\begin{equation}
={\bf K}_3[\o{{\bf u}}] + \left(\frac{n}{2} + 2\right)\e{\bf K}_2[\o{{\bf u}}]
+ \frac{1}{2}\left(\frac{n}{2} + 1\right)\left(\frac{n}{2} + 2\right)\e^2\o{{\bf u}}_x. \nonumber
\label{eq:ut3}
\end{equation}

Let us define two one-parameter linear transformations of independent
variables
\begin{equation}
\left(
\begin{array}{c}
\o{x}\\
\o{t}_2
\end{array}
\right) = B_2(\e)
\left(
\begin{array}{c}
x\\
t_2
\end{array}
\right) =
\left(
\begin{array}{cc}
1&\left(\frac{n}{2} + 1\right)\e\\[0.2cm]
0&1
\end{array}
\right)
\left(
\begin{array}{c}
x\\
t_2
\end{array}
\right)
\label{eq:tr1}
\end{equation}
and
\begin{equation}
\left(
\begin{array}{c}
\o{x}\\
\o{t}_2\\
\o{t}_3
\end{array}
\right) = B_3(\e)
\left(
\begin{array}{c}
x\\
t_2\\
t_3
\end{array}
\right) =
\left(
\begin{array}{ccc}
1&\left(\frac{n}{2} + 1\right)\e&
\frac{1}{2}\left(\frac{n}{2} + 1\right)\left(\frac{n}{2} + 2\right)\e^2\\[0.2cm]
0&1&\left(\frac{n}{2} + 2\right)\e\\[0.2cm]
0&0&1
\end{array}
\right)
\left(
\begin{array}{c}
x\\
t_2\\
t_3
\end{array}
\right)
\label{eq:tr2}
\end{equation}
It is easy to check that matrices $B_2(\e)$ and $B_3(\e)$ satisfy group
relation $B(\e)B(\e_1) = B(\e + \e_1)$.
By straightforward calculations, it can be verified that:

(i) system of evolution equation
${\bf u}_{t_2} = {\bf K}_2[{\bf u}] = \Lambda[{\bf u}]{\bf u}_x$
is invariant under transformations
(\ref{eq:6}) and (\ref{eq:tr1}),

(ii) a pair of systems
${\bf u}_{t_2} = {\bf K}_2[{\bf u}] = \Lambda[{\bf u}]{\bf u}_x$ and
${\bf u}_{t_3} = {\bf K}_3[{\bf u}] = \Lambda^2[{\bf u}]{\bf u}_x$
is invariant under transformations
(\ref{eq:6}) and (\ref{eq:tr2}).

We can undertake to generalize transformations (\ref{eq:tr1}) and
(\ref{eq:tr2}). Let us denote ${\bf t} = (x, t_2, t_3,..., t_N)^T$. We suppose
that matrix $B_N(\e)$ defining transformation of independent variables
\begin{equation}
\o{{\bf t}} = B_N(\e){\bf t}
\label{eq:indv}
\end{equation}
is upper diagonal matrix with units in main
diagonal and is written in the form
$$
B_N(\e) =
\left(
\begin{array}{cccc}
1&b_{2,1}\e&...&b_{N,1}\e^{N-1}\\[0.2cm]
0&1&\ddots&\vdots\\[0.2cm]
\vdots&\ddots&\ddots&b_{N,N-1}\e\\[0.2cm]
0&...&0&1
\end{array}
\right).
$$
Here $b_{k,l}$ are some coefficients to be determined.
Necessary condition in order that (\ref{eq:indv}) complete transformation
(\ref{eq:6}) to define one-parameter group is the relation
$$
B_N(\e)B_N(\e_1) = B_N(\e + \e_1),
$$
which element-wise can be written as
\begin{equation}
\begin{array}{c}
\displaystyle b_{m+r,m}\left(\e + \e_1\right)^r = b_{m+r,m}\left(\e^r +
\sum_{i=1}^{r-1}C_r^i\e^i\e_1^{r-i} + \e_1^r\right) \\[0.3cm]
\displaystyle =b_{m+r,m}\e^r + \sum_{i=1}^{r-1}b_{m+r,m+i}b_{m+i,m}\e^i\e_1^{r-i} +
b_{m+r,m}\e_1^r.
\end{array}
\label{eq:24}
\end{equation}
Thus the relation
\begin{equation}
C_r^ib_{r+k,k} = b_{r+k,k+i}b_{k+i,k}
\label{eq:25}
\end{equation}
must be fulfilled for each $i = 1,..., r-1$.
The solution of Eq. (\ref{eq:25}) is uniquely given by
$$
b_{m+r,m} = \frac{1}{r!}\prod_{i=1}^rb_{k+i,k+i-1}.
$$
Thus to derive coefficients $b_{ml}$ there is a need to calculate $b_{m,m-1}$.
In order to do that we observe that
\begin{equation}
\o{{\bf u}}_{t_m} = {\bf K}_m[\o{{\bf u}}] +
b_{m,m-1}\e{\bf K}_{m-1}[\o{{\bf u}}] +
O(\e^2).
\label{eq:Oe2}
\end{equation}
Taking into account (\ref{eq:f1}) we have
$$
\o{{\bf u}}_{t_m} = \Lambda^{m-1}[\o{{\bf u}}]\o{{\bf u}}_{x} +
(m-1)\e\Lambda^{m-2}[\o{{\bf u}}]\o{{\bf u}}_{x} +
O(\e^2)
$$
$$
= \Lambda^{m-2}[\o{{\bf u}}]\left({\bf K}_2[\o{{\bf u}}] +
\frac{n}{2}\e \o{{\bf u}}_{x}\right) +
(m-1)\e\Lambda^{m-2}[\o{{\bf u}}]\o{{\bf u}}_{x} + O(\e^2)
$$
$$
= \Lambda^{m-2}[\o{{\bf u}}]{\bf K}_2[\o{{\bf u}}] +
\left(\frac{n}{2} + m -1\right)\e\Lambda^{m-2}[\o{{\bf u}}]\o{{\bf u}}_{x} +
O(\e^2)
$$
\begin{equation}
= {\bf K}_m[\o{{\bf u}}] + \left(\frac{n}{2} + m -1\right)\e {\bf K}_{m-1}
[\o{{\bf u}}] + O(\e^2).
\label{eq:21}
\end{equation}
Here we also consider that $\Lambda[\o{{\bf u}}]{\bf K}_2[\o{{\bf u}}] =
{\bf K}_3[\o{{\bf u}}] + O(\e^2)$. From (\ref{eq:21}) it follows that
$$
b_{m+r,m} =
\frac{1}{r!}\prod_{i=1}^r\left(\frac{n}{2} + m + i - 1\right).
$$

In what follows, we will denote the groups of one-parameter transformations
(\ref{eq:7}), (\ref{eq:indv}) by ${\cal G}_{N,n}$. The infinitesimal generators
(vector fields) of ${\cal G}_{N,n}$ can be calculated immediately to yield
$$
{\bf v}_{N,n} = \sum_{i=1}^{N-1}\left(\frac{n}{2} + i\right)t_{i+1}\frac{\partial}
{\partial t_i} - \sum_{j=1}^nju_{j+1}\frac{\partial}{\partial u_j},
$$
where $u_{n+1}=1$. For convenience, here we denote $t_1 = x$. As can be checked,
respective characteristics are given by
$$
Q_{N,n}({\bf u}) = -\sum_{i=1}^{N-1}\left(\frac{n}{2} + i\right)t_{i+1}{\bf u}_{t_i} -
A^{\prime}(0){\bf u} - {\bf d}^{\prime}(0),
$$
where $A^{\prime}(0)$ and ${\bf d}^{\prime}(0)$ denote a valuation of
derivatives with respect to group parameter calculated at $\epsilon = 0$,
being $A^{\prime}(0) = \sum_{i=1}^{n-1}iE_{i+1,i}$ and
$$
{\bf d}^{\prime}(0) =
\left(
\begin{array}{c}
0\\
\vdots\\
0\\
n
\end{array}
\right).
$$
Here $E_{ij}$ denotes matrix with unit in $(i,j)$ place and zeros elsewhere.

Repeatedly applying recursion operator $\Lambda[{\bf u}]$ to $Q_{N,n}$, we
obtain infinite
sequence of characteristics $Q_{N,n}^{(i)}$. In particular, once applying
$\Lambda[{\bf u}]$ to $Q_{N,n}$ yields the characteristic
$$
Q_{N,n}^{(1)}({\bf u}) = -\sum_{i=1}^{N}\left(\frac{n}{2} + i - 1\right)t_{i}{\bf u}_{t_i}
- J{\bf u},
$$
where $J = {\rm diag}(n, n-1,..., 1)$.
This characteristic corresponds to scaling invariance group. Below we
construct scaling invariance transformations making use a transformation property
of recursion operator $\Lambda[{\bf u}]$.
\vspace{0.4cm}

\leftline{{\bf B. Scaling symmetry transformations}}

Let us investigate invariance properties of the systems (\ref{eq:hier})
w.r.t. scaling transformations. We denote
$p_1 = \frac{1}{4}\partial_x^2 + u_1 + \frac{1}{2}u_{1x}\partial_x^{-1}$,
$p_i = u_i + \frac{1}{2}u_{ix}\partial_x^{-1},\;\;i = 2,..., n$. It is
obvious that relation
$$
\left(
\begin{array}{cccc}
\beta^{n+1}&0&...&0\\[0.3cm]
0&\beta^n&\ddots&\vdots\\[0.3cm]
\vdots&\ddots&\ddots&0\\[0.3cm]
0&...&0&\beta^2
\end{array}
\right)
\left(
\begin{array}{ccccc}
0&0&...&0&p_1\\[0.3cm]
-1&0&...&0&p_2\\[0.3cm]
0&-1&\ddots&\vdots&p_3\\[0.3cm]
\vdots&\ddots&\ddots&0&\vdots\\[0.3cm]
0&...&0&-1&p_n
\end{array}
\right)
$$
\begin{equation}
= \left(
\begin{array}{ccccc}
0&0&...&0&\beta^n p_1\\[0.3cm]
-1&0&...&0&\beta^{n-1}p_2\\[0.3cm]
0&-1&\ddots&\vdots&\beta^{n-2}p_3\\[0.3cm]
\vdots&\ddots&\ddots&0&\vdots\\[0.3cm]
0&...&0&-1&\beta p_n
\end{array}
\right)
\left(
\begin{array}{cccc}
\beta^{n}&0&...&0\\[0.3cm]
0&\beta^{n-1}&\ddots&\vdots\\[0.3cm]
\vdots&\ddots&\ddots&0\\[0.3cm]
0&...&0&\beta
\end{array}
\right),
\label{eq:beta}
\end{equation}
where $\beta$ is arbitrary number, is valid. In what follows, $\beta$
will be group parameter.

Let us introduce scaling transformation of variables $x$ and $u_i$:
$$
x = \beta^{n/2}\o{x},\;\; u_i = \beta^{i-n-1}\o{u}_i.
$$
Consequently Eq. (\ref{eq:beta}) we comes to the relation
\begin{equation}
\left.\Lambda[{\bf u}]\right|_{{\bf u} = A(\beta)\o{{\bf u}}} =
A(\beta)\left(\beta\Lambda[\o{{\bf u}}]\right)A^{-1}(\beta),
\label{eq:rel1}
\end{equation}
where $A(\beta) = {\rm diag}(\beta^n,..., \beta)$. Using the Eq.
(\ref{eq:rel1}), we obtain equation
$$
\o{{\bf u}}_{t_m} = \beta^{(n/2) + m -1}
\Lambda^{m-1}[\o{{\bf u}}]\o{{\bf u}}_{\o{x}}
$$
from which it follows that by scaling evolution parameter $t_m$ as
$\o{t}_m = \beta^{(n/2) + m -1}t_m$ we arrive at
$$
\o{{\bf u}}_{\o{t}_m} = \Lambda^{m-1}[\o{{\bf u}}]\o{{\bf u}}_{\o{x}}
$$
Thus we have the following statement:

{\bf Corollary 1:} {\it The system of equations
$$
{\bf u}_{t_m} = \Lambda^{m-1}[{\bf u}]{\bf u}_{x}
$$
with the recursion operator (\ref{eq:ro}) is invariant under
scaling transformation
\begin{equation}
\o{x} = \beta^{n/2}x,\;\;
\o{t}_m = \beta^{(n/2) + m - 1}t_m,\;\;
\o{u}_i = \beta^{i-n-1}u_i.
\label{eq:sctr}
\end{equation}
}

One can assign, to each variable, scaling dimension $[\cdot]$, so that
$[x] = n/2$, $[t_m] = (n/2) + m - 1$ and $[u_i] = i - n - 1$.

{\large
\leftline{{\bf III. INVARIANT SOLUTIONS}}
}

Given Galilei-like symmetry group ${\cal G}_{N,n}$, corresponding one-parameter
family of solutions of (\ref{eq:hier}) is given by
\begin{equation}
{\bf u}(\e, x, t_2,..., t_N) = A(\e){\bf u}(\o{x}, \o{t}_2,..., \o{t}_N) +
{\bf d}(\e)
\label{one-parameter}
\end{equation}
Next we are going, using standard symmetry method, to construct
${\cal G}_{N,n}$-invariant solutions of equations (\ref{eq:hier}). To do
this, we need to find the set of generators of ${\cal G}_{N,n}$-invariant
functions algebra.

{\bf Proposition 1:} {\it For the group ${\cal G}_{N,n}$, there exist
functionally independent differential invariants in the form:
\begin{equation}
\begin{array}{l}
\displaystyle
T_j = t_j + \sum_{k=1}^{N-j-1}(-1)^k\alpha_{j,k}
\frac{t_{j+k}t_{N-1}^k}{t_N^k},\;\; j = 1,..., N-2,\\
T_N = t_N,\\
\displaystyle
v_j = u_j + \sum_{k=1}^{n-j}(-1)^k\beta_{j,k}u_{j+k}u_n^k,\;\; j = 1,..., n-1,\\
\displaystyle
v_n = u_n + \frac{2nt_{N-1}}{(2N+n-2)t_N},
\end{array}
\label{diff-inv}
\end{equation}
where the coefficients $\alpha_{j,k}$ and $\beta_{j,k}$ are positive
rational numbers to be determined.}

{\it Proof.} It is evident that $T_N$ is an invariant.
Consider $T_j$, for $j = 1,..., N-3$. We have
$$
{\bf v}_{N,n}(T_j) =
\left(\frac{n}{2} + j\right)t_{j+1}
$$
$$
- \left(\frac{n}{2} + j + 1\right)\alpha_{j,1}\frac{t_{j+2}t_{N-1}}{t_N} +...+
(-1)^{N-j-1}\left(\frac{n}{2} + N - 1\right)\alpha_{j,N-j-1}
\frac{t_Nt_{N-1}^{N-j-1}}{t_N^{N-j-1}}
$$
$$
+ \left(\frac{n}{2} + N - 1\right)t_N\left\{-\alpha_{j,1}\frac{t_{j+1}}{t_N}
\right.
$$
$$
\left.
+ \alpha_{j,2}\frac{2t_{j+2}t_{N-1}}{t_N^2} +...
+ (-1)^{N-j-1}\alpha_{j,N-j-1}
\frac{(N-j-1)t_{N-1}^{N-j-1}}{t_{N}^{N-j-1}}
\right\}.
$$
Require that ${\bf v}_{N,n}(T_j) = 0$. Collecting similar terms and equating it
to zero, we obtain
recurrence relations:
$$
t_{j+1}\;\; : \;\; \left(\frac{n}{2} + j\right) =
\left(\frac{n}{2} + N - 1\right)\alpha_{j,1},
$$
$$
\frac{t_{j+k}t_{N-1}^{k-1}}{t_{N}^{k-1}}\;\; : \;\;
\left(\frac{n}{2} + j + k - 1\right)\alpha_{j,k-1} =
k\left(\frac{n}{2} + N - 1\right)\alpha_{j,k},\;\; k = 2,..., N - j - 2,
$$
$$
\frac{t_{N-1}^{N-j-1}}{t_{N}^{N-j-2}}\;\; : \;\;
\left(\frac{n}{2} + N - 2\right)\alpha_{j,N-j-2} =
(N-j)\left(\frac{n}{2} + N - 1\right)\alpha_{j,N-j-1}.
$$
Separately, consider the case $j = N - 2$. Assuming that ${\bf v}_{N,n}(T_{N-2}) = 0$,
we obtain relation
$$
{\bf v}_{N,n}(T_{N-2}) = \left(\frac{n}{2} + N - 2\right)t_{N-1} - 2\alpha_{N-2,1}
\left(\frac{n}{2} + N - 1\right)t_{N-1} = 0
$$
from which it follows
$$
2\left(\frac{n}{2} + N - 1\right)\alpha_{N-2,1} = \left(\frac{n}{2} + N - 2\right).
$$

So, all coefficients $\alpha_{j,k}$ are uniqely defined by recurrence relations above.
>From these relations, we easy obtain
$$
\alpha_{j,k} = \frac{1}{k!\left(\frac{n}{2} + N - 1\right)^k}
\prod_{i=1}^k\left(\frac{n}{2} + i + j - 1\right),\;\; j = 1,..., N-j-2,
$$
$$
\alpha_{j,N-j-1} = \frac{(N-j-1)}{(N-j)(N-j-1)!\left(\frac{n}{2} + N - 1\right)^{N-j-1}}
\prod_{i=1}^{N-j-1}\left(\frac{n}{2} + i + j - 1\right).
$$

By simple calculations, it is easy to prove that $v_n$ is an invariant of
${\cal G}_{N,n}$. Consider $v_j$, for $j = 1,..., n-1$. Requiring that
${\bf v}_{N,n}(v_j) = 0$ we obtain
$$
{\bf v}_{N,n}(v_j) = -n\left\{-\beta_{j,1}u_{j+1} + 2\beta_{j,2}u_{j+2}u_n +...+
(-1)^{n-j-1}(n-j-1)\beta_{j,n-j-1}u_{n-1}u_n^{n-j-2}
\right.
$$
$$
\left.
+ (-1)^{n-j}(n-j)\beta_{j,n-j}u_n^{n-j}
\right\} - ju_{j+1} + (j+1)\beta_{j,1}u_{j+2}u_n - (j+2)\beta_{j,2}u_{j+3}u_n^2 +...
$$
\vspace{0.1cm}
$$
- (-1)^{n-j-1}(n-1)\beta_{j,n-j-1}u_n^{n-j} - (-1)^{n-j}n\beta_{j,n-j}u_n^{n-j} = 0.
$$
Collecting similar terms we obtain recurrence relations:
$$
u_{j+1}\;\; : \;\; n\beta_{j,1} = j,
$$
$$
u_{j+k}u_n^k\;\; : \;\; kn\beta_{j,k} = (j+k-1)\beta_{j,k-1},\;\; k = 2,..., n-j-1,
$$
\vspace{0.1cm}
$$
u_n^{n-j}\;\; : \;\; (n-j)n\beta_{j,n-j} = (n-1)\beta_{j,n-j-1} - n\beta_{j,n-j},
$$
from which it follows that
$$
\beta_{j,k} = \frac{C^k_{j+k-1}}{n^k},\;\;\; k = 1,..., n-j-1
$$
$$
\beta_{j,n-j} = \frac{n-j}{n-j+1}\frac{C^{n-j}_{n-1}}{n^{n-j}}.
$$
It is obvious that all invariants determined above are functionally independent
and generate the algebra of ${\cal G}_{N,n}$-invariant functions. $\Box$

It should be noted that $[T_j] = [t_j]$ and $[v_j] = [u_j]$.
Denote ${\bf T} = (T_1=X, T_2,..., T_{N-2})^T$ and ${\bf t} = (x, t_2,...,
t_{N-2})^T$. In what follows, it will be useful following two lemmas.

{\bf Lemma 2:} {\it The relation
\begin{equation}
{\bf T} = B_{N-2}(\tau){\bf t} + {\bf c}(\tau),
\label{simple}
\end{equation}
where
$$
\tau = -\frac{t_{N-1}}{\left(\frac{n}{2} + N - 1\right)t_N}
$$
and ${\bf c}(\tau) = (c_1(\tau),..., c_{N-2}(\tau))^T$, where
$$
c_i(\tau) = \frac{t_{N-1}\tau^{N-j-1}}
{(N-j)(N-j-2)!}
\prod_{k=1}^{N-j-1}
\left(\frac{n}{2} + j + k - 1\right)
$$
is fulfilled.
}

{\it Proof:} It is easy to prove the validity of the relation (\ref{simple})
by simple straightforward calculations. $\Box$

{\bf Lemma 3:} {\it The matrix $A(\tau)$ and vector ${\bf d}(\tau)$ as functions
of $\tau$ satisfy differential equations:}
\begin{equation}
A^{-1}(\tau)A^{\prime}(\tau) = A^{\prime}(0),\;\;\;
A^{-1}(\tau){\bf d}^{\prime}(\tau) = {\bf d}^{\prime}(\tau).
\label{diff-eq}
\end{equation}

{\it Proof:} Taking into account group identities $A(\tau + \tau_1) =
A(\tau)A(\tau_1)$ and ${\bf d}(\tau + \tau_1) = A(\tau){\bf d}(\tau_1) +
{\bf d}(\tau)$, we easy obtain
$$
A^{\prime}(\tau) =
\lim_{\tau_1\rightarrow 0}
\frac{A(\tau + \tau_1) - A(\tau)}{\tau_1} =
A(\tau)\lim_{\tau_1\rightarrow 0}
\frac{A(\tau_1) - I}{\tau_1} = A(\tau)A^{\prime}(0),
$$
$$
{\bf d}^{\prime}(\tau) =  \lim_{\tau_1\rightarrow 0}
\frac{{\bf d}(\tau + \tau_1) - {\bf d}(\tau)}{\tau_1} = A(\tau)
\lim_{\tau_1\rightarrow 0}
\frac{{\bf d}(\tau_1)}{\tau_1} = A(\tau){\bf d}^{\prime}(0)
$$
>From these relations, it follows Eq. (\ref{diff-eq}). $\Box$

Now, using differential invariants (\ref{diff-inv}), we can construct
invariant solutions. By turns, we define more convenient variables $U_j$ by
$$
U_n = v_n,\;\;\; U_j = v_j - \sum_{k=1}^{n-j}(-1)^k\beta_{j,k}U_{j+k}U_n^k,\;\;\;
j = n-1,..., 1.
$$
Observe that
$$
u_n = U_n + n\tau = U_n + C_n^{n-1}\tau.
$$
Next
$$
u_{n-1} =
v_{n-1} + \beta_{n-1,n}u_n^2 = v_{n-1} +
\beta_{n-1,n}\left(U_n^2 + 2n\tau U_n + n^2\tau^2\right)
$$
$$
= U_{n-1} + 2n\beta_{n,n-1}\tau U_n + n^2\beta_{n,n-1}\tau^2.
$$
Using the relations
$$
2n\beta_{n,n-1} = C_{n-1}^{n-2},\;\;\; n^2\beta_{n,n-1} = C_n^{n-2}
$$
we obtain
$$
u_{n-1} = U_{n-1} + C_{n-1}^{n-2}\tau U_n + C_n^{n-2}.
$$
Thus, it is naturally to conjecture that the relation
\begin{equation}
u_j = U_j + \sum_{k=1}^{n-i}C_{i+k-1}^{i-1}\tau^kU_{i+k} + C_n^{i-1}\tau^{n-i-1}
\label{u_j}
\end{equation}
holds for $i = 1,..., n$. We observe that, consequently (\ref{bf}) and
(\ref{aik}), the relation (\ref{u_j}) can be rewritten in vector form as
$$
{\bf u} = A(\tau){\bf U} + {\bf d}(\tau).
$$
Thus, we comes to the following ansatz for ${\cal G}_{N,n}$-invariant solutions
\begin{equation}
{\bf u}(x, t_2,..., t_n) = A(\tau){\bf U}(X, T_2,..., T_{N-2}, T_N) +
{\bf d}(\tau).
\label{ansatz}
\end{equation}
In what follows, we will consider the case $N\ge 3$.
Taking into account Eq. (\ref{one-parameter}) and lemma 2, we obtain

{\bf Proposition 2:} {\it Substitution of the ansatz (\ref{ansatz})
into the equations (\ref{eq:hier}), for $m = 1,..., N-2$, yields}
\begin{equation}
{\bf U}_{T_{m}} = {\bf K}_{m}[{\bf U}],\;\;\;
m = 1,..., N-2.
\label{""}
\end{equation}

Here ${\bf K}_{m}[{\bf U}]$ are vector fields ${\bf K}_{m}[{\bf u}]$, where
variables $u_{ix}^{(k)}$ are replaced by $U_{iX}^{(k)}$.

{\bf Proposition 3:} {\it Substitution of the ansatz (\ref{ansatz})
into the system
$$
{\bf u}_{t_{N-1}} = {\bf K}_{N-1}[{\bf u}]
$$
by virtue (\ref{""}), yields following constraint:}
\begin{equation}
\left(\frac{n}{2} + N - 1\right)T_N{\bf K}_{N-1}[{\bf U}] +
\sum_{i=1}^{N-3}\left(\frac{n}{2} + i\right)T_{i+1}{\bf K}_i[{\bf U}] +
A^{\prime}(0){\bf U} + {\bf d}^{\prime}(0)= 0
\label{constraint}
\end{equation}

{\it Proof:} We have
$$
{\bf u}_{t_{N-1}} = - \frac{1}{\left(\frac{n}{2} + N -1\right)t_N}
A^{\prime}(\tau){\bf U} - \frac{1}{\left(\frac{n}{2} + N -1\right)t_N}
{\bf d}^{\prime}(\tau)
$$
$$
+ A(\tau)\left\{{\bf U}_{X}
\left( - \alpha_{1,1}\frac{t_2}{t_N}
+ 2\alpha_{1,2}\frac{t_3t_{N-1}}{t_N^2} -... +
(-1)^{N-3}(N-3)\alpha_{1,N-3}\frac{t_{N-2}t_{N-1}^{N-4}}{t_N^{N-3}}
\right.\right.
$$
$$
\left.
+ (-1)^{N-2}(N-1)\alpha_{1,N-2}\frac{t_{N-1}^{N-2}}{t_N^{N-2}}\right)
+ {\bf U}_{T_2}\left( - \alpha_{2,1}\frac{t_3}{t_N}
+ 2\alpha_{2,2}\frac{t_4t_{N-1}}{t_N^2} -...
\right.
$$
$$
\left.
+ (-1)^{N-4}(N-4)\alpha_{2,N-4}\frac{t_{N-2}t_{N-1}^{N-5}}{t_N^{N-4}}
+  (-1)^{N-3}(N-2)\alpha_{2,N-3}\frac{t_{N-1}^{N-3}}{t_N^{N-3}}\right) +...
$$
$$
\left.
+ {\bf U}_{T_{N-2}}\left(
- 2\alpha_{N-2,1}\frac{t_{N-1}}{t_N}
\right)
\right\} =
A(\tau)\left\{{\bf K}_{N-1}[{\bf U}] + b_{N-1,N-2}\tau{\bf K}_{N-2}[{\bf U}] +...
+ b_{N-1,1}\tau^{N-2}{\bf U}_{X}\right\}.
$$
Taking into account Eq. (\ref{""}), lemma 3 and the relations
\begin{equation}
\begin{array}{c}
\displaystyle
\left(\frac{n}{2} + N - 1\right)\alpha_{i,1} =
\left(\frac{n}{2} + i\right),\\[0.3cm]
\displaystyle
k\left(\frac{n}{2} + N - 1\right)\alpha_{i,k} =
\left(\frac{n}{2} + i\right)\alpha_{i+1,k-1},\\[0.3cm]
\displaystyle
(N-i)\left(\frac{n}{2} + N - 1\right)\alpha_{i,N-i-1} =
\left(\frac{n}{2} + i\right)\alpha_{i+1,N-i-2} +
\frac{b_{N-1,i}}{\left(\frac{n}{2} + N - 1\right)^{N-i-2}}
\end{array}
\label{the}
\end{equation}
for $k = 2,..., N-i-2,\;\; i = 1,..., N-2$,
we comes to (\ref{constraint}). $\Box$

{\bf Proposition 4:} {\it Substitution of the ansatz (\ref{ansatz})
into the system
$$
{\bf u}_{t_N} = {\bf K}_N[{\bf u}]
$$
by virtue Eqs. (\ref{""}) and (\ref{constraint}), yields}
\begin{equation}
{\bf U}_{T_N} = {\bf K}_N[{\bf U}]
\label{UTN}
\end{equation}

{\it Proof:} We have
$$
{\bf u}_{t_N} =  \frac{t_{N-1}}{\left(\frac{n}{2} + N -1\right)t_N^2}
A^{\prime}(\tau){\bf U} + \frac{t_{N-1}}{\left(\frac{n}{2} + N -1\right)t_N^2}
{\bf d}^{\prime}(\tau)
$$
$$
+ A(\tau)\left\{{\bf U}_{X}
\left( \alpha_{1,1}\frac{t_2t_{N-1}}{t_N^2}
- 2\alpha_{1,2}\frac{t_3t_{N-1}^2}{t_N^3} +...
-(-1)^{N-3}(N-3)\alpha_{1,N-3}\frac{t_{N-2}t_{N-1}^{N-3}}{t_N^{N-2}}
\right.\right.
$$
$$
\left.
- (-1)^{N-2}(N-2)\alpha_{1,N-2}\frac{t_{N-1}^{N-1}}{t_N^{N-1}}\right)
+ {\bf U}_{T_2}\left(  \alpha_{2,1}\frac{t_3t_{N-1}}{t_N^2}
- 2\alpha_{2,2}\frac{t_4t_{N-1}^2}{t_N^3} +...
\right.
$$
$$
\left.
- (-1)^{N-4}(N-4)\alpha_{2,N-4}\frac{t_{N-2}t_{N-1}^{N-4}}{t_N^{N-3}}
-  (-1)^{N-3}(N-3)\alpha_{2,N-3}\frac{t_{N-1}^{N-2}}{t_N^{N-2}}\right) +...
$$
$$
\left.
+ {\bf U}_{T_{N-2}}\left(
 \alpha_{N-2,1}\frac{t_{N-1}^2}{t_N^2}
\right)
+ U_{T_N}
\right\} =
A(\tau)\left\{{\bf K}_{N}[{\bf U}] + b_{N,N-1}\tau{\bf K}_{N-1}[{\bf U}] +...
+ b_{N,1}\tau^{N-1}{\bf U}_{X}\right\}.
$$
Now taking into account Eq. (\ref{""}), lemma 3, the relations (\ref{the})
and the relation
$$
(N-i-1)\left(\frac{n}{2} + N - 1\right)\alpha_{i,N-i-1} =
\left(\frac{n}{2} + i\right)\alpha_{i+1,N-i-2} +
\frac{b_{N,i}}{\left(\frac{n}{2} + N - 1\right)^{N-i-1}}
$$
we obtain (\ref{UTN}). $\Box$

We observe that constraint (\ref{constraint}) is Galilean self-similarity
condition
\begin{equation}
- Q_{N,n}({\bf U}) =
\sum_{i=1}^{N-1}\left(\frac{n}{2} + i\right)T_{i+1}
{\bf K}_i[{\bf U}] + A^{\prime}(0){\bf U} + {\bf d}^{\prime}(0) = 0,
\label{galilean}
\end{equation}
where $T_{N-1} = 0$. Consequence of (\ref{galilean}) is scaling self-similarity
condition
\begin{equation}
-Q_{N,n}^{(1)}({\bf U}) = \sum_{i=1}^{N}\left(\frac{n}{2} + i - 1\right)T_{i}{\bf K}_i[{\bf U}] + J{\bf U} = 0,
\label{scaling}
\end{equation}
where $T_{N-1}=0$. By virtue (\ref{""}) and (\ref{UTN}), we can replace in Eq.
(\ref{scaling})
$$
{\bf K}_i[{\bf U}] \rightarrow {\bf U}_{T_i},
$$
for $i = 1,..., N,\; i\neq N-1$, to obtain the constraint
$$
\left(\frac{n}{2} + N - 1\right)T_{N}{\bf U}_{T_N} +
\sum_{i=1}^{N-2}\left(\frac{n}{2} + i - 1\right)T_{i}{\bf U}_{T_i} + J{\bf U} = 0.
$$

{\it Example.} Separately consider the case $N = 3$. The ansatz
$$
{\bf u}(x, t_2, t_3) = A(\tau){\bf U}(X, T_3) + {\bf d}(\tau),\;\;
\tau = - \frac{t_2}{\left(\frac{n}{2} + 2\right)t_3},\;\;
X = x - \frac{(n+2)t_2^2}{2(n+4)t_3},\;\;
T_3 = t_3
$$
gives
\begin{equation}
\left(\frac{n}{2} + 2\right)T_3{\bf K}_2[{\bf U}] + A^{\prime}(0){\bf U} +
{\bf d}^{\prime}(0) = 0,
\label{constr1}
\end{equation}
\begin{equation}
\left(\frac{n}{2} + 2\right)T_3{\bf U}_{T_3} + \frac{n}{2}X{\bf U}_X +
J{\bf U} = 0.
\label{constr2}
\end{equation}
More explicitly (\ref{constr1}) is given by
\begin{equation}
\frac{5}{2}T_3
\left(\frac{1}{4}U_{XXX} + \frac{3}{2}UU_X\right) + 1 = 0
\label{by1}
\end{equation}
for $n=1$ and
\begin{equation}
\begin{array}{c}
\displaystyle
\left(\frac{n}{2} + 2\right)T_3
\left(\frac{1}{4}U_{nXXX} + U_1U_{nX} + \frac{1}{2}U_nU_{1X}\right) + U_2 = 0,\\[0.3cm]
\displaystyle
\left(\frac{n}{2} + 2\right)T_3
\left(-U_{i-1,X} + U_iU_{nX} + \frac{1}{2}U_nU_{iX}\right) + iU_{i+1} = 0,\;\;
i = 2,..., n-1
\\[0.3cm]
\displaystyle
\left(\frac{n}{2} + 2\right)T_3
\left(-U_{n-1,X} + \frac{3}{2}U_nU_{nX}\right) + n = 0
\end{array}
\label{by2}
\end{equation}
for $n\ge 2$. Eq. (\ref{constr2}), which presents scaling self-similar
constraint can be immediately solved to yield
\begin{equation}
U_i(X, T_3) = \frac{1}{T_3^p}f_i(Z),\;\;\;
Z = \frac{X}{T_3^q}
\label{solved}
\end{equation}
where
$$
p = \frac{2(n-i+1)}{n+4},\;\;\;
q = \frac{n}{n+4}.
$$
Introducing (\ref{solved}) into (\ref{by1}) and (\ref{by2}) we gives,
respectively,
\begin{equation}
\frac{5}{2}
\left(\frac{1}{4}f_{ZZZ} + \frac{3}{2}ff_Z\right) + 1 = 0
\label{res1}
\end{equation}
for $n=1$ and
\begin{equation}
\begin{array}{c}
\displaystyle
\left(\frac{n}{2} + 2\right)
\left(\frac{1}{4}f_{nZZZ} + f_1f_{nZ} + \frac{1}{2}f_nf_{1Z}\right) + f_2 = 0,\\[0.3cm]
\displaystyle
\left(\frac{n}{2} + 2\right)
\left(-f_{i-1,Z} + f_if_{nZ} + \frac{1}{2}f_nf_{iZ}\right) + if_{i+1} = 0,\;\;
i = 2,..., n-1
\\[0.3cm]
\displaystyle
\left(\frac{n}{2} + 2\right)
\left(-f_{n-1,Z} + \frac{3}{2}f_nf_{nZ}\right) + n = 0.
\end{array}
\label{res2}
\end{equation}
for $n\ge 2$.

According to Ablowitz---Ramani---Segur conjecture$^8$, equations (\ref{res1})
and (\ref{res2}) should have the Painlev\'e property.
Single integration of Eq. (\ref{res1}) gives first Painlev\'e trancedent$^9$:
$$
f^{\prime\prime} + \frac{6}{5}f^2 + \frac{8}{5}Z = 0
$$
whose canonical form is $f^{\prime\prime} = 6f^2 + Z$. The system (\ref{res2})
in the case $n=2$ is reduced to second Painlev\'e trancedent. Putting
$$
f_1 = \frac{3}{4}f^2 + \frac{2}{3}Z,\;\;\;
f_2 = f
$$
and once integrating, we obtain second Painlev\'e equation:
$$
f^{\prime\prime} + 2f^3 + \frac{8}{3}Zf + \alpha = 0
$$
whose canonical form is $f^{\prime\prime} = 2f^3 + Zf + \alpha$.

The problem to be addressed: to investigate Painlev\'e property for the
system (\ref{res2}), for every $n$.
\newpage
\noindent
$^1$ L.V. Ovsiannikov, {\it Group Analysis of Differential Equations},
(Academic, New York, 1982).\\[0.4cm]
$^2$ N.H. Ibragimov, {\it Transformation Groups Applied to Mathematical
Physics}, (Reidel, Dordrecht, 1985)\\[0.4cm]
$^3$ P.J. Olver, {\it Applications of Lie Groups to Differential Equations},
Graduate texts in Mathematics (Springer-Verlag, New York, 1986).\\[0.4cm]
$^4$ Y. Kosmann-Schwarzbach, Lett. Math. Phys. {\bf 38}, 421 (1996).\\[0.4cm]
$^5$ B. Fuchssteiner, Nonlinear Anal. Theor. Meth. Appl. {\bf 3}, 849 (1979).\\[0.4cm]
$^6$ L.M. Alonso, J. Math. Phys. {\bf 21}, 2342 (1980).\\[0.4cm]
$^7$ A.K. Svinin, J. Phys. A: Math. Gen. {\bf 32}, 5499 (1999).\\[0.4cm]
$^8$ M.J. Ablowitz, A. Ramani, H.J. Segur, J. Math. Phys. {\bf 21}, 715 (1980).\\[0.4cm]
$^9$ E.L Ince, {\it Ordinary Differential Equations}, (Dower, New York, 1956).
\end{document}